# Proceedings to the 24rd International Workshop "What Comes Beyond the Standard Models", July 5 -- July 11, 2021, Bled, Slovenia, [Virtual Workshop --- July 5.--11. 2021]

Scientific Committee:

- *John Ellis* (King's College London/CERN)
- *Roman Jackiw* (MIT)
- *Masao Ninomiya* (Yukawa Institute for Theoretical Physics, Kyoto University)

Editors: N.S. Mankoč Borštnik, H.B. Nielsen, D. Lukman and A. Kleppe

Bled Workshops in Physics Vol. 22, No.1
DMFA Založnistvo, Ljubljana, December 2021

The Members of the Organizing Committee of the International Workshop "What Comes Beyond the Standard Models", Bled, Slovenia, state that the articles published in The Proeedings to the 24th International Workshop "What Comes Beyond the Standard Models", Bled, Slovenia are refereed at the Workshop in intense in-depth discussions.

The whole Proceedings in pdf format can be found also in the Proceedings and in the Virtual Institute of Astroparticle Physics (CosmoVia)

# Contents:

## Preface in English Language

The series of annual workshops on "What Comes Beyond the Standard Models?" started in 1998 with the idea of Norma and Holger for organizing a real workshop, in which participants would spend most of the time in discussions, confronting different approaches and ideas. All this time we have been looking for answers to the question of what the laws of nature are. And we learned a lot. This year in July the 24th workshop took place.
Workshops have always taken place in the picturesque town of Bled by the lake of the same name, surrounded by beautiful mountains and offering pleasant walks and mountaineering. Except for the last two years, 2020 and 2021, when workshop has again taken place in July, but without personal conversations all day and late at night, even between very relaxing walks and mountaineering due to COVID-19 pandemic. We have, however, a very long tradition of videoconferences (cosmovia), enabling discussions and explanations with laboratories all over the world. This enabled us to have these two years a virtual workshop, resembling Bled workshops as much as possible.

In our very open minded, friendly, cooperative, long, tough and demanding discussions several physicists and even some mathematicians have contributed. Most of topics that have been presented and discussed in our Bled workshops deal withthe proposals for explaining physics beyond the so far accepted and experimentally confirmed both standard models — in physics of fermion and boson fields and cosmology — in order to understand the origin of assumptions of both standard models and be consequently able to propose new theories, models and to make predictions for future experiments.

Although in all these years most of participants were theoretical physicists, many of them with their own suggestions how to make the next step beyond the accepted models and theories, experts from experimental laboratories were and are very appreciated, helping a lot to understand what do measurements really tell and which kind of predictions can best be tested. Also in the last two years we tried to keep our habit of (long) presentations (with breaks and continuations over several days), followed by very detailed discussions. The authors of the articles worked hard and with enthusiasm already before the presentations and as well when preparing the articles for this Proceedings in such a short time.

However, as lectures and especially discussions over the Internet are more exhausting than live, many issues remain open, unresolved, also undefined and undicussed. And we did not succeed to continue the discussions over the Internet after the workshop, even though we tried, because of several reasons, one of them was that the computer of one of the organizers broke down. Here are some questions that we have not really discussed yet but have just started discussing: How efficient are models offering a small next step beyond both standard models, suggesting experiments which could test such a model, to be able to explain all the observations so far, or at least many of them? There are several contributions with such proposals presented in this Proceedings, most of them trying to explain what does the dark matter consist of. Would the confrontation of these models with string theories, for example, or with the spin-charge-family theory, which offers the explanation for all the assumptions of both standard models, offering as well the explanation for several phenomena observed so far, with the dark matter and the matter/antimatter asymmetry included, the theory is presented in this Proceedings, help to understand our universe better and also to easier propose relevant experiments having correspondingly more chance to be the right next step beyond both standard models?

Combining knowledge, ideas and hard work could increase our opportunities to recognize the real next step beyond the two standard models and to suggest trustable experiments. In particular if experimentalists would be involved in discussions. Experiments are expensive. Although the black holes are experimentally well confirmed objects, the quantum mechanics of black holes is not really known. This knowledge is needed for heavy black holes as well as if we accept the possibility that the space-time is larger than $(3 + 1)$, as it is in string theories, in Kaluza-Klein theories and in the spin-charge-family theory (with fermions interacting with gravity only) with space-time (13 + 1) or larger (appearing in two contributions in this paper), or might be even infinite (since zero and infinite are easy to be accepted, all other possibilities need the explanation). Do we understand what in this context the primordial black holes, discussed in this and last year Proceedings, mean and do they appear before or after the electroweak phase transition? Is in the time of the formation of the primordial black holes space-time already (3 + 1)? What happens inside the primordial black holes and what happens within the very massive experimentally confirmed black holes? Can string theories within M-theory help to understand the quantum gravity even in the context that the internal space of fermions and bosons describe the Clifford algebra objects? If Nature does use the Clifford algebra objects to describe internal space of fermions and bosons, what explains the second quantization postulates for fermions and bosons, as explained in one contribution in this Proceedings, can the quantum mechanics of black holes be easier understood? Would the novel string theory, discussed in this Proceedings, where noninteracting objects representing strings, are themselves bound states of strings and might explain bound states of objects with the high symmetry, be helpful as well for describing the heavy black hole objects? Even under the assumption that the internal space of fermion and gravitational fields are described with the Clifford algebra objects? It does happen (after having a vision and after a hard work) that a new way of treating quantum mechanics of bound systems, this time the superconformal quantum mechanics and light-front holography used in hadron physics, presented in this Proceedings, opens a new understanding of dynamics and symmetries of bound states. To understand the start and the starting expansion of our universe the knowledge of quantum gravity and the knowledge of the internal space of fermions and bosons is needed.
Some periods and some phenomena in the expansion of our universe can be explained in the context of string theories, as it is the period of the inflation described with one article in this Proceedings. When has the inflation taken place and how is it connected with the today non observed extra dimensions? Do we have besides the ordinary matter also domains of antimatter in our universe? What are properties of the antimatter? Do domains of antimatter contain mostly the dark matter? What is the interaction of matter with antimatter on the border of both domains? In the spin-charge-family theory the laws and the interactions are the same — for fermions, antifermions and dark matter. In several talks these problems were discussed, some of them presented only as a talk on the website on Proceedings and in the Virtual Institute of Astroparticle Physics (CosmoVia)

Symmetries play the essential role on all levels of physics, on the level of elementary fermion and boson fields, of cosmology and also of matter of all kinds. In the theories assuming more than (3 + 1) dimensions with fermions which interact with gravity only, like there are the Kaluza-Klein theories, the spin-charge-family theory is also of this kind, as well as string theories, the symmetry origins in the Lorentz invariance of spacetime, manifesting also in the internal space of boson and fermion fields. At observable (that is low) energies the Lorentz symmetry of higher dimensions manifests in (3 + 1) spacetime (after breaking the starting symmetry of spacetime) the symmetry of the internal space of fermion and boson fields only, which usually is described by the group theoretical methods. The symmetries of elementary fermion and boson fields are discussed in several contributions talks of Bled 2021 workshop, manifesting that all these different understanding of symmetries have a strong overlap. Some talks about symmetries appear in these Proceedings, the others can

only be found on the follow-up page of the official website of the Workshop:
Proceedings and on the Cosmovia Forum
There are several other topics, discussed in this Proceedings, like
i. What is indeed the origin of masses of fermions and gauge fields?
ii. What modes of gravitational waves can be observed?
iii. How far can we interpret experiments correctly if we accept the standard model only?
iv. The DAMA/LIBRA experiment, measuring the collisions of the dark matter particles with the ordinary matter, reports on the newest results of the annual modulation of data together with the measurements, collected in more than 10years. v. Choosing the action for the assumed laws of Nature to predict experiments one must be able to calculate properties of systems accurately enough. One must be able to evaluate the renormalizability, the anomalies, for any proposed theory. The reader can find some answers to these questions in this Proceedings.

This year neither the cosmological nor the particle physics experiments offered much new, as also has not happened in the last three years, which would offer new insight into the elementary fermion and boson fields and also into cosmological events, although a lot of work and effort have been put in.

However, there are more and more cosmological evidences, which require the new step beyond both standard models, the one of the elementary fermion and boson fields and of cosmology. The understanding the universe through the cosmological theories and theories of the elementary fermion and boson fields have, namely, so far never been so dependent on common knowledge and experiments in both fields.

Although cosmovia served the discussions all the time (and we are very glad that we did have in spite of pandemic also the 24th workshop), it was not like previous workshops. Discussions were fiery and sharp, at least during some talks.

But this was not our Bled workshop. Effective discussions require the personal presence of the debaters, as well as of the rest of participants, which interrupt the presentations with questions all the time.

And let us add also this year that due to the on line presentations we have students participants, who otherwise would not be able to attend the Bled conference, the travel expenses are too high for them. The organizers hope that the virus will be defeated at least up to next year, although the data are not supporting our hope. Let our hope be valid for all over the world, especially for the young generation, as well as for the Bled Workshop 2022, so that we will in July next year meet at Bled.

Since, as every year, also this year there has been not enough time to mature the discussions into the written contributions, only two months, authors can not really polish their contributions. Organizers hope that this is well compensated with fresh contents.

The reader can find all the talks and soon also the whole Proceedings on the official http://bsm.fmf.uni-lj.si/bled2021bsm/presentations.html website of the Workshop: Proceedings and on the Cosmovia Forum

The organizers are thanking Dragan Lukman for his excellent technical support to more then twenty years of Bled workshops, entitled "What comes beyond the standard models", in particular for his excellent work done on Proceedings. In July of this year we learned how small the step is between to be and not to be. The member of editors Dragan Lukman, our friend and the man who recognized clearly the essential problems of our planet, is not among us any longer. He left us after this year workshop due to the hart attack. We are missing him very much, also during the preparation of The Proceedings, although we copied his way of preparing the Proceedings, using his styles. In memory of Dragan, we added a short summary of his work and Astri's song, which says a lot about Dragan.The organizing committee thanks Astri Kleppe, who offered to take over Dragan's work on the Proceedings when our hardship was greatest. The organizing committee thanks also Ana Bračič and Anamarija Borštnik Bračič who have done the translations of English abstracts to the Slovenian language. Let us conclude this preface with a heartfelt and warm thank to all the participants, present via videoconference, for their lectures and especially for the very prolific discussions and, nevertheless, an excellent atmosphere. We are very sorry that some of participants could not prepare their talks as contributions to the Proceedings.

Norma Mankoč Borštnik, Holger Bech Nielsen, Maxim Y. Khlopov, (the Organizing comittee)
Norma Mankoč Borštnik, Holger Bech Nielsen, Dragan Lukman, Astri Kleppe (the Editors)
Ana Bračič, Anamarija Bračič Borštnik (the translators into Slovenian)
Ljubljana, December 2021

# Predgovor (Preface in Slovenian Language)

Vsakoletne delavnice z naslovom ,,Kako preseči oba standardna modela, kozmološkega in elektrošibkega" ("What Comes Beyond the Standard Models?") sta postavila leta 1998 Norma in Holger z namenom, da bi udeleženci v izčrpnih diskusijah kritično soočali različne ideje in teorije. V vsem tem času smo iskali odgovore na vprasanje kakšni so zakoni narave. In se veliko naučili.
To leto je stekla 24. delavnica.
Delavnice domujejo v Plemljevi hiši na Bledu ob slikovitem jezeru, kjer prijetni sprehodi in pohodi na čudovite gore, ki kipijo nad mestom, ponujajo priložnosti in vzpodbudo za diskusije. Tako je bilo vse do zadnjih dveh let.
Tudi zadnji dve leti, v letu 2020 in 2021, sta bili delavnici v juliju, vendar nam je tokrat covid-19 onemogočil srečanje v Plemljevi hiši. Tudi diskutirali nismo med hojo okoli jezera ali med hribolazenjem. Vendar nam je dolgoletna iskušnja s "cosmovio" — videopovezavami z laboratoriji po svetu — omogočila, da je tudi letos stekla Blejska delavnica, tokrat prek interneta.
K našim zelo odprtim, prijateljskim, dolgim in zahtevnim diskusijam, polnim iskrivega sodelovanja, je prispevalo veliko fizikov in celo nekaj matematikov. V večini predavanj in razprav so udeleženci poskusili razumeti in pojasniti predpostavke obeh standadnih modelov, elektrošibkega in barvnega v fiziki osnovnih delcev in polj ter kozmološkega, predpostavke in napovedi obeh modelov pa vskladiti z meritvami in opazovanji, da bi poiskali model, ki preseže oba standardna modela, kar bi omogočilo zanesljivejše napovedi za nove poskuse.
Čeprav je večina udeležencev teoretičnih fizikov, mnogi z lastnimi idejami kako narediti naslednji korak onkraj sprejetih modelov in teorij, so še posebej dobrodošli predstavniki eksperimentalnih laboratorijev, ki nam pomagajo v odprtih diskusijah razjasniti resnično sporočilo meritev in nam pomagajo razumeti kakšne napovedi so potrebne, da jih lahko s poskusi dovolj zanesljivo preverijo.
Tudi v zadnjih dveh letih smo poskušali ohraniti navado, da so bile predstavitve dolge, ker so jih udeleženci prekinjali z vprašanji, da bi bili privzetki in predpostavke jasni. Predavanja so se zato po dveh urah prekinila in se nadaljevala naslednje dni.

Avtorji prispevkov so trdo in z navdušenjem delali, da so pripravili predavanja, in da so v tako kratkem času pripravili članke za ta zbornik. Ker pa so predavanja preko interneta bolj naporna kot v predavanja v živo, so mnoga vprašanja ostala odprta, nerazjasnena, tudi nedefinirana in nerešena.Ni nam uspelo nadaljevati pogovorov preko interneta po končani delavnici, četudi smo poskušali. Razlogi so bili različni, med njimi sesutje računalnika ene(ga) od organizatorjev.

Med vprašanji, ki smo jih odprli, pa o njih nismo uspeli zares razpravljati, so: Kako učinkoviti so lahko modeli, ki ponudijo majhen naslednji korak glede na oba standardna modela, da bi nato predlagali izvedbo poskusov, ki naj povedo ali so taki modeli v skladu z naravo, pri iskanju odgovorov na vsa odprta vprašanja, ali vsaj na del odprtih vprašanj? Kar nekaj prispevkov v tem zborniku, ki poskušajo pojasniti, iz česa utegne biti temna snov, je te vrste.

Ali bi bilo smiselno in bi zmogli primerjati te predloge, denimo, s teorijami strun ali s teorijo spinov, nabojev, družin, ki že odgovori na odprta vprašanja obeh stan- dardnih modelov in ponudi tudi napovedi, ki jih je potrebno preveriti, za temno snov in tudi za druga kozmološka opaženja.

Združevanje znanja, idej in vloženega dela bi lahko povečalo možnosti, da pre- poznamo, kaj je pravi naslednji korak, ki prinaša odgovore na odprta vprašanja v fiziki osnovnih fermionskih in bozonskih polj in kozmologiji ter bi pomagalo predlagati zaupanja vredne poskuse, ki bodo domneve potrdili, posebej, če bi pri diskusijah tvorno sodelovali tudi experimentalci. Experimenti so dragi.

Čeprav so črne luknje eksperimentalno potrjeni objekti, kvantna mehanika črnih lukenj v resnici ni znana. Vendar je to znanje potrebno, če sprejmemo možnost, da je prostor-čas več kot $(3 + 1)$-razsežen, kot to domnevajo teorije strun in Kaluza-Kleinove teorije, da je njegova razsežnost morda celo $(13 + 1)$ ali več, kot domneva teorija spin-charge-family (s fermioni, ki interagirajo samo z gravitacijsko silo, z dvema prispevkoma v tem zborniku), ali kot domneva tudi teorija strun, ali pa je lahko neskončen, saj je nič in neskončno enostavno sprejeti, vse druge možnosti potrebujejo pojasnila. Kako v tem kontekstu razumeti primordialne črne luknje? Ali se pojavijo po elektrošibkem faznem prehodu? Ali nastanejo že prej? Ali tedaj prostor-čas že učinkuje kot $(3 + 1)$-razsežen? In kaj se dogaja znotraj teh primordialnih črnih lukenj? Kaj pa se dogaja znotraj zelo masivnih črnih lukenj? Ali lahko teorije strun v kontekstu M-teorije pomagajo razumeti kvantno gravitacijo tudi, če notranji prostor fermionov in bozonov določa Cliffordova algebra? Ali bi teorija, imenovana nova teorja strun, poročilo je najti v tem zborniku, s strunami iz inertnih objektov, ki so dejansko strune vezane v struno z veliko stopnjo simetrije, bila sprejemljiva tudi, če bi notranje stopnje fermionov in bozonov določali Cliffordovi objekti? Bi bile take strune koristne tudi za razumevanje kvantne mehanike zelo masivnih črnih lukenj?

Zgodi se, da nov način obravnavanja kvantne mehanike vezanih sistemov, kakršen jeuporaba superkonformne kvantne mehanike in holografije "light-front" v hadron-ski fiziki, omogoči nov pogled in novo razumevanje dinamike in simetrij vezanih stanj. Poročilo o tem prinaša zbornik.

Za razumevanje nastanka in začetne širitve našega vesolja sta kvantna gravitacija in poznavanje notranjega prostora fermionov in bozonov potrebno orodje. Vsaj nekatera obdobja širitve, obdobje inflacije denimo, je mogoče razložiti v kontekstu teorije strun, o čemer poriča en prispevek.Ali imamo v našem vesolju poleg običajne snovi tudi domene antisnovi? Je anti- snov pretežno iz temne snovi? Kakšne so interakcije snovi s temno snovjo? O tem diskutirajo avtorji nekaterih prispevkov v tem zborniku.

Teorija spina, nabojev in družin gradi na predpostavki, da so zakoni gibanja enotni — za snov, za antisnov in za temno snov.

Simetrije igrajo bistveno vlogo na vseh ravneh fizike: v kozmologiji, v fiziki osnovnih fermionskih in bozonskih polj, tudi v fiziki vseh vrst snovi. V teorijah, ki predpostavijo da ima prostor več kot $(3+1)$ razsežnost, in da interagirajo fermioni samo z gravitacijskimi bozoni — take so Kaluza-Kleinove teorije, tudi teorija spina-nabojev-družin, pa tudi teorije strun — je izvor simetrije v Lorentzovi invariantnosti prostor-časa, ki vključuje tudi notranji prostor fermionov in bozonov. Pri opaženih (nizkih) energijah (po zlomitvi začetne simetrije) določa lastnosti prostora z razsežnostimi $d > (3+1)$ notranji prostor fermionskih in bozonskih polj, kar opazimo v $d = (3+1)$-razsežnem prostor-času kot simetrije, ki jih lahko opišemo tudi z metodami teorije grup.

Simetrije osnovnih fermionskih in bozonskih polj so obravnavane v nekaj prispevkih. Koliko skupnega imajo različni pristopi pa bi bilo potrebno in koristno raziskati.

Naj omenimo še nekatere druge teme, k jih prispevki v zborniku obravnavajo: i. Kaj je pravi vzrok, da imajo fermioni in nekatera bozonska polja maso?

ii. Kako dolgo še lahko pravilno interpretiramo rezultate poskusov z uporabo samo standardnega modela?

iii. Experiment DAMA/LIBRA prinša poročilo o zadnjih rezultatih meritev letne modulacije trkov delcev temne snovi z običajno snovjo v njihovih merilnih aparaturah, povzema pa tudi vse dolgoletne meritve.

iv. Ko izberemo model, moramo v modelu znati primerjati rezultate meritev dovolj natančno. Moramo vedeti ali je teorija renormalizabilna, ali ima anomalije, in kako se računov lotiti. Tudi na taka vprašanja poskuša odgovoriti eden od prispevkov. Tako kot v preteklih treh letih tudi to leto niso eksperimenti v kozmologiji in fiziki osnovih fermionskih in bozonskih polj ponudili rezultatov, ki bi omogočili nov vpogled v fiziko osnovnih delcev in polj, čeprav je bilo vanje vloženega veliko truda.

Vse več je tudi kozmoloških meritev, za katere se zdi, da jih standardni model osnovnih fermionskih in bozonskih polj ne more pojasniti in vse bolj kozmološke meritve in opažanja ter experimentalne meritve v fiziki osnovnih fermionskih in bozonskih polj določajo iskanje teorije, ki lahko pojasni vse predpostavke standardnega modela, pa tudi vsa nova kozmološka opažanja in vse nove meritve ter predlaga prave experimente. Četudi je cosmovia poskrbela, da so diskusije tekle ves čas, tako kot je bilo na vseh delavnicah doslej, blejskih diskusij v živo diskusije po internetu niso mogle nadomestiti. Diskusije so bile ognjevite in ostre, vsaj pri nekaterih predavanjih, vendar potrebujejo učinkovite diskusije osebno prisotnost diskutantov in poslušalcev, ki z vprašanji poskrbijo, da je debata razumljiva vsem. Tudi študentom internet ne more nadomestiti dobrega učitelja.

Organizatorji upamo, da bo vsaj do naslednjega leta virus premagan, četudi ta trenutek naše upanje ni podprto s statističnimi podatki. Naj naše upanje velja zaves svet, za mlado generacijo pa še posebej, pa tudi za Blejsko delavnico 2022, da bo stekla v živo na Bledu.

Ker je vsako leto le malo časa od delavnice do zaključka redakcije, manj kot dva meseca, avtorji ne morejo dovolj skrbno pripraviti svojih prispevkov, vendar upamo, da to nadomesti svežina prispevkov.

Bralec najde zapise vseh predavanj in kmalu tudi letošnji zbornik na uradnem naslovu Delavnice na medmrežju:

[Proceedings](#) in na [Cosmovia Forum](#)

Zahvaljujemo se Draganu Lukmanu za odlično tehnično podporo več kot dva-jsetletnim blejskim delavnicam z naslovom "Kako preseči oba standardna modela", ter za tehnično pripravo zbornikov. Letos smo izvedeli, kako majhen je korak med biti in ne biti. Član uredniškega odbora Dragan Lukman, naš prijatelj in človek, ki je jasno prepoznaval probleme naše družbe, ni več med nami. Zapustil nas je kmalu za tem, ko se je končala letošnja Blejska delavnica. Imel je srčni napad.

Pogrešamo ga, še posebej zdaj med pripravo zbornika, čeprav nam je zapustil tehnično znanje priprave zbornika..

Draganu v spomin smo dodali kratek povzetek njegovega dela ter Astrino pesem, ki veliko pove o Draganu.

Organizacijski odbor se zahvaljuje Astri Kleppe, ki se je ponudila, da prevzame Draganovo delo na zborniku, ko je bila naša stiska največja.

Zahvaljujemo se tudi Ani Bračič in Anamariji Borštnik Bračič za prevode angleškega teksta v slovenščino. Naj zaključimo ta predgovor s prisrčno in toplo zahvalo vsem udeležencem, prisotnim preko videokonference, za njihova predavanja in še posebno za zelo plodne diskusije in kljub vsemu odlično vzdušje.

Zelo nam je žal, da nekateri udeleěnci niso utegnili pripraviti poleg predavamj tudi zapis teh predavanj v obliki prispevkov.

Norma Mankoč Borštnik, Holger Bech Nielsen, Maxim Y. Khlopov, (Organizacijski odbor)
Norma Mankoč Borštnik, Holger Bech Nielsen, Dragan Lukman, Astri Kleppe (uredniki)
Ana Bračič, Anamarija Bračič Borštnik (prevodi v slovenščino)
Ljubljana, grudna (decembra) 2021

# Invited Talks

The Proceedings of 24th Bled Workshop "What Comes beyond the Standard Models", 2021
Title: Type IIB moduli stabilisation, inflation and waterfall fields
Authors: I. Antoniadis, O. Lacombe G. K. Leontaris
Comments: 16 pages, published in The Proceedings of XXIV Bled Workshop "What Comes Beyond the Standard Model?" Bled, July 5-11 2021
Subjects: physics.gen-th

The Proceedings of 24th Bled Workshop "What Comes beyond the Standard Models", 2021
Title: New and recent results, and perspectives from DAMA/LIBRA–phase2
Authors: R. Bernabei, P. Belli, A. Bussolotti, V. Caracciolo, R. Cerulli, N. Ferrari, A. Leoncini, V. Merlo, F. Montecchia, F. Cappella , A. d'Angelo , A. Incicchitti, A. Mattei, C.J. Dai, X.H. Ma, X.D. Sheng, Z.P. Ye
Comments: 19 pages, published in The Proceedings of XXIV Bled Workshop "What Comes Beyond the Standard Model?" Bled, July 5-11 2021
Subjects: physics.gen-exp
arXiv: 2110.13011

Title: The multicomponent dark matter structure and its possible observed manifestations
Authors: V. Beylin, V. Kuksa, M. Bezuglov, D. Sopin
Comments: 16 pages, published in The Proceedings of XXIV Bled Workshop "What Comes Beyond the Standard Model?" Bled, July 5-11 2021
Subjects: High Energy Physics - Phenomenology (hep-ph)
arXiv: 2111.09042

Title: Numerical simulation of Bohr-like and Thomson-like dark atoms with nuclei
Authors: T. E. Bikbaev, M. Yu. Khlopov, A. G. Mayorov
Comments: Published in The Proceedings of XXIV Bled Workshop "What Comes Beyond the Standard Model?" Bled, July 5-11 2021
Subjects: High Energy Physics - Phenomenology (hep-ph)
arXiv: 2112.02453 [pdf, ps, other]
Title: Supersymmetric and Other Novel Features of Hadron Physics from Light-Front Holography
Authors: S. J. Brodsky
Comments: ReportNumber SLAC-PUB-17634, 38 pages, Contribution to The Proceedings of the XXIV Bled Workshop "What Comes Beyond the Standard Models?" Bled, July 5-11 2021
Subjects: General Physics (physics.gen-ph);High Energy Physics - Phenomenology (hep-ph)
arXiv: 2111.00139 [pdf, ps, other]
Title: Charge asymmetry of new stable quarks in baryon asymmetrical Universe
Authors: A. Chaudhuri, M. Yu. Khlopov
Comments: 11 pages, 5 figures, Contribution to The Proceedings of the XXIV Bled Workshop "What Comes Beyond the Standard Models?" Bled, July 5-11 2021
Subjects: General Physics (physics.gen-ph);High Energy Physics - Phenomenology (hep-ph)
arXiv: 2111.00139 [pdf, ps, other]
Title: Entopy release in Electroweak Phase Transition in 2HDM
Authors: A. Chaudhuri, M. Yu. Khlopov, S. Porey
Comments: 11 pages, 5 figures, Contribution to The Proceedings of the XXIV Bled Workshop "What Comes Beyond the Standard Models?" Bled, July 5-11 2021
Subjects: General Physics (physics.gen-ph);High Energy Physics - Phenomenology (hep-ph)
arXiv: 2111.07704 [pdf, ps, other]
Title: Gravitational waves in the modified gravity
Authors: S. R. Chowdhury, M. Yu. Khlopov
Comments: 11 pages, Contribution to The Proceedings of the XXIV Bled Workshop "What Comes Beyond the Standard Models?" Bled, July 5-11 2021
Subjects: General Relativity and Quantum Cosmology (gr-qc)
The Proceedings of 24th Bled Workshop "What Comes beyond the Standard Models", 2021
Title: Representing rational numbers and divergent geometric series by binary graphs
Authors: E. Dmitrieff
Comments: 16 pages, published in The Proceedings of XXIV Bled Workshop "What Comes Beyond the Standard Model?" Bled, July 5-11 2021
Subjects: physics.gen-ph
The Proceedings of 24th Bled Workshop "What Comes beyond the Standard Models", 2021
Title: Neutrino masses within a SU(3) family symmetry and a 3+5 scenario
Authors: A. Hernandez-Galeana
Comments: 16 pages, published in The Proceedings of XXIV Bled Workshop "What Comes Beyond the Standard Model?" Bled, July 5-11 2021
Subjects: physics.gen-th
arXiv: 2111.14114 [pdf, ps, other]

Title: Statistical analyses of antimatter domains, created by nonhomogeneous baryosynthesis in a baryon asymmetrical Universe
Authors: M. Yu. Khlopov, O. M. Lecian
Comments: 11 pages, Contribution to The Proceedings of the XXIV Bled Workshop "What Comes Beyond the Standard Models?" Bled, July 5-11 2021
Subjects: physics.gen-ph
arXiv: 2112.03375 [pdf, ps, other]

Title: BSM Cosmology from BSM Physics
Authors: M. Yu. Khlopov
Comments: 11 pages, Contribution to The Proceedings of the XXIV Bled Workshop "What Comes Beyond the Standard Models?" Bled, July 5-11 2021
Subjects: High Energy Physics - Phenomenology (hep-ph); Cosmology and Nongalactic Astrophysics (astro-ph.CO); High Energy Astrophysical Phenomena (astro-ph.HE)
arXiv: 2112.00361 [pdf, ps, other]

Title: Researching of magnetic cutoff for local sources of charged particles in the halo of the Galaxy
Authors: A. O. Kirichenko, M. Yu. Khlopov, A. V. Kravtsova, A. G. Mayorov
Comments: 8 pages, Contribution to The Proceedings of the XXIV Bled Workshop "What Comes Beyond the Standard Models?" Bled, July 5-11 2021
Subjects: High Energy Astrophysical Phenomena (astro-ph.HE)
arXiv: 2201.13236 [pdf, ps, other]

Title: Mass as a dynamical quantity
Authors: M. Land
Comments: 13 pages, published in The Proceedings of XXIV Bled Workshop "What Comes Beyond the Standard Model?" Bled, July 5-11 2021
Subjects: physics.gen-ph
arXiv: 2112.03375 [pdf, ps, other]

Title: How do Clifford algebras show the way to the second quantized fermions with unified spins, charges and families, and to the corresponding second quantized vector and scalar gauge field
Authors: N.S. Mankoc Borstnik
Comments: Published in The Proceedings to 24th Bled Workshop "What Comes beyond the Standard Models", 2021, Zaloznistvo DMFA, Ljubljana, December 2021
Subjects: General Physics (physics.gen-ph); High Energy Physics - Theory (hep-th)
arXiv: arxiv: 2012.09640,[v2] [pdf, ps, other]

Title: The achievements of the spin-charge-family theory so far
Authors: N.S. Mankoc Borstnik
Comments: Published in Proceedings to 24th Bled Workshop "What Comes beyond the Standard Models", 2021, Zaloznistvo DMFA, Ljubljana, December 2021
Subjects: General Physics (physics.gen-ph); High Energy Physics - Theory (hep-th)
arXiv: 2111.10879v1 [pdf, ps, other]

Title: Atomic Size Dark Matter Pearls, Electron Signal
Authors: Holger Bech Nielsen, C. D. Froggatt
Comments: 22 pages, Published in Proceedings to 24th Bled Workshop "What Comes beyond the Standard Models", 2021, Zaloznistvo DMFA, Ljubljana, December 2021
Subjects: [hep-ph]
arXiv: 2111.05106v1 [pdf, ps, other]

Title: Novel String Field Theory and Bound State, Projective Line, and sharply 3-transitive group

Authors: Holger Bech Nielsen, Masao Ninomiya
Comments: 22 pages, contribution to The Proceeding of XXIV Bled Workshop "What Comes Beyond the Standard Model?" Bled, July 5-11 2021
Subjects: General physics, phenomenological (physics.gen-ph)
arXiv: 2110.00366 [pdf, ps, other]
Title: Galactic model with a phase transition from dark matter to dark energy
Authors: I. Nikitin
Comments: 22 pages, contribution to The Proceeding of XXIV Bled Workshop "What Comes Beyond the Standard Model?" Bled, July 5-11 2021
Subjects: General physics
arXiv: 2111.13318 [pdf, ps, other]
Title: Ultraviolet divergences in supersymmetric theories regularized by higher derivatives
Authors: K. Stepanyantz
Comments: 20 pages, contribution to The Proceeding of XXIV Bled Workshop "What Comes Beyond the Standard Model?" Bled, July 5-11 2021
Subjects: High Energy Physics – Theory (hep-th)

*Because of the pandemic, the Bled Workshop has now been virtual for the two last years, 2020 and 2021. Not all the talks come as articles in this year's Proceedings, but all the talks can be found on the official website of the Workshop and on the Cosmovia forum:* The Proceedings of 24th Bled Workshop "What Comes beyond the Standard Models", 2021 *and* Cosmovia

A. Addazi: The multicomponent dark matter structure and its possible observed manifestations.

L. Bonora: HS Yang-Mills-like models.

S. Kabana: Sexaquarks, the unexpected Dark Matter candidate.

E. Kiritsis: Coleman de Luccia transitions, and their implications for Quantum Field Theories in De Sitter space.

R. Mohapatra: The Next Symmetry of Nature.

Q. Shafi: Quest for Unification.

# The Platform of Virtual Institute of Astroparticle Physics for Studies of BSM Physics and Cosmology (by M.Yu. Khlopov)

The Virtual Institute of Astroparticle Physics (VIA) which operates on website http://viavca.in2p3.fr/site.html, has provided the platform for our online virtual meetings. Since 2014 VIA online lectures combined with individual work on Forum acquired the form of Open Online Courses. Aimed to individual work with students the Course is not Massive, but the account for the number of visits to VIA site converts VIA in a specific tool for MOOC activity. VIA sessions, being a traditional part of Bled Workshops' program, have con- verted at XXIV Bled Workshop "What comes beyond the Standard models?" into the only format, challenging to preserve the creative nonformal atmosphere of meetings in Bled, Slovenia. We openly discuss the state of art of VIA platform: http://bsm.fmf.uni-lj.si/bled2021bsm/presentations.html https://bit.ly/bled2021bsm.

[M. Yu. Khlopov](): *Challenging BSM physics and cosmology on the online platform of Virtual Institute of Astroparticle physics*,
[The Proceedings to 24th Bled Workshop "What Comes beyond the Standard Models", 2021, Zaloznistvo DMFA, Ljubljana, December 2021]()

# Postscriptum

*In Memory of Mag. Dragan Lukman, 11 March 1962 - 19 July 2021*

Words do not obey thoughts and feelings when writing in memory of someone who has been a collaborator for many years, a friend I talked to over short coffees about all open questions of our world: in physics, cosmology, mathematics, about society, human life, about values; we just never talked about personal life. Dragan Lukman joined us in Koper on my Project on elementary fermion and boson fields (the project at the Department of Physics, Faculty of Mathematics and Physics, University of Ljubljana), when I managed to establish the Institute of Technical and Natural Sciences in Koper (PINT). He also was involved in common projects with industry. At that time, the Bled workshop was held for the second year. Dragan took over the technical side of editing workshops and Proceedings of the workshops up to this year 2021, the 24th workshop. He was all the time an excellent helper and a good friend to all.
The first research in the field of physics of elementary fermionic and bosonic fields, in which Dragan participated, were at first published in the Proceedings of the workshops "What comes beyond the standard models". They belong to a project entitled spin-charge-family theory, which I am developing since 1992, also together with colleagues and students. There are still some articles that are not yet prepared for publication in international journals in which Dragan participated.
An overview of all Proceedings can be found on the home page of the Bled Workshops [Proceedings]() and after 2008 also on the [Cosmovia Forum]()
Proceedings are cited in articles published also in international journals in this field, among them those coauthored with Dragan.
A song can say a lot and Astri Kleppe on behalf of all of us, who appreciated Dragan and liked him, wrote the poem, appearing in this Proceedings.
Norma Susana Mankoč Borštnik norma.mankoc@fmf.uni-lj.si

Dragan Lukman was introduced into the research work in the middle of eighties of previous century when Slovenia started with the "1000 Young Researchers Project". After finishing his diploma work at the Department of Physics Dragan decided to expand his field of interest to the field of mathematics. He enrolled the postgraduate course at the Department of Mathematics and simultaneously he participated in the research work at the National Institute of Chemistry as a member of the Laboratory for Molecular Modelling. In due course he accomplished all the necessary steps to attain the degree of master of mathematical sciences. His participation in the scientific work resulted in ten publications in international scientific journals. Dragan was able to cope with the research work in quite diverse fields such as strict statistical mechanics, the application of molecular dynamics simulation of biological systems and even technologically oriented studies of mechanical properties of fullerenes.
Prof. dr. Branko Borštnik, The head of The Laboratory for Molecular Modelling at the National Institute for Chemistry Ljubljana, Slovenia in the period when Dragan Lukman was member of the group

Mag. Dragan Lukman, holding Master of Science degree in Mathematics and Bachelor degree in Physics, both degrees received from University of Ljubljana, has approached me, after important recommendations from Prof.dr. Norma Mankoč Borštnik, in May 2019 with an interest to apply for a research position in my research project Quantum Localization in Chaotic Systems being carried out at CAMTP - Center for Applied Mathematics and Theoretical Physics of the University of Maribor, funded by the Slovenian Research Agency ARRS. In our first interview with him it was immediately obvious that he has quite wide experience in working with various research groups in Slovenia, predominantly with Norma Mankoč and her coworkers, but also with others, with broad knowledge in physics and mathematics, and in computational physics. Therefore my decision to offer him the job was easy. Thus he has joined my core research group, a part of CAMTP, whose members also are dr. Qian Wang, dr. Črt Lozej (my PhD student at the time) and dr. Benjamin Batistić (also my former PhD student, 2015, now postdoc). We started to work together with Dragan on 1 June 2019. Our main object of study was the phenomenon of quantum or dynamical localization in classically chaotic systems, one of the central issues in the domain of quantum chaos. More precisely, we have been studying very extensively and deeply the localization phenomena in the so called lemon billiards, a special family of two dimensional billiards with extremely rich behaviour both classically and quantaly.
They are important paradigmatic model systems. The selection of billiards was made possible only thanks to the extensive calculations by Črt Lozej in the courseof his PhD thesis. Dragan started his work quite enthusiastically, and was using mainly the software codes developed over the many years by Benjamin Batistić and recently very drastically improved and expanded by Črt Lozej. In doing so we were discovering very many exciting results which emerged by our heavy computations, and Dragan was always very careful, fast, responsive and reliable, with good physical insight, presenting the results in a shortest possible time, working every day from early morning until the late afternoon, and even on weekends at home. Based on the results under his cooperation four important papers have been produced, 3 of them already published in excellent journals (Physical Review E, Physics MDPI, Nonlinear Phenomena in Complex Systems), the fourth one just in the progress of writing. Therefore Dragan's contribution to our results is quite essential and appreciated. Dragan was a very pleasant personality, highly modest and quiet person, always helpful, never complaining, and deeply dedicated to his work, not only at our institute, but also in other groups. We did not know much about his personal life, as he was a very shy person and did not show emotions, but this does not mean that he was not sensible and empathic. The tragic news about his sudden death on his way to work in the early morning on Monday 19 July 2021 was a great shock for all of us. We shall remember him as a wonderful fellow and a very good researcher. Our papers with him are a long lasting remembrance of him.
Prof.dr. Marko Robnik, member of EASA Founder and Director of CAMTP - Center for Applied Mathematics and Theoretical Physics, University of Maribor
Robnik@uni-mb.si

*Draganu Lukmanu v spomin in zahvalo, 11. marec 1962 - 19. julij 2021*

Prave besede kar ne stečejo v zapis v spomin nekomu, ki je bil dolga leta sodelavec, prijatelj in s katerim sva ob kratkih kavicah prediskutirala vsa odprta vprašanja tega sveta, v fiziki, v kozmologiji, v matematiki, v družbi, v človekovem življenju, o vrednotah, le o osebnem življenju nisva govorila nikoli. Dragan Lukman se mi je pridružil v Kopru na projektu Fizike osnovih delcev in polj, Oddelka za fiziko Fakultete za matematiko in fiziko Univerze v Ljubljani, ko mi je uspelo ustanoviti Primorski inštitut za naravoslovne in tehnične vede Koper. Sodeloval je tudi na projektih, ki smo jih razvili z gospodarstvom.
Tedaj je Blejska delavnica tekla že drugo leto. Prevzel je tehnično plat urejanja delavnice in zbornika delavnice vse do letošnje 24. delavnice. Bil je vseskozi izvrsten pomočnik in dober prijatelj vsem.
Prve raziskave na področju fizike osnovnih fermionskih in bozonskih polj, pri katerih je Dragan sodeloval, so bile najprej objavljene v zbornikih delavnic "What comes beyond the standard models". Sodijo v projekt z naslovom spin-charge- family theory, ki ga razvijam, tudi skupaj s sodelavci in študenti, že od leta 1992. Je še nekaj prispevkov, ki še niso dozoreli za objavo v mednarodnih revijah, pri katerih je Dragan sodeloval.
Pregled vseh zbornikov je najti na domači strani Blejskih delavnic [Proceedings](#) po letu 2008 pa tudi na [Cosmovia Forum](#)
V prispevkih zbornikov so citirani članki, ki so, potem ko so dozoreli, objavljeni v mednarodnih revijah s tega področja, tudi tisti v soavtorstvu z Draganom.
Pesem pove lahko zelo veliko in Astri Kleppe je v imenu vseh nas, ki smo Dragana cenili in imeli radi, napisala pesem, ki jo objavljamo v tem zborniku.
Norma Susana Mankoč Borštnik
norma.mankoc@fmf.uni-lj.si

Dragan Lukman se je pridružil moji raziskovalni skupini v Laboratoriju za molekularno modeliranje na Nacionalnem Kemijskem inštitu sredi osemdesetih let prejšnjega stoletja, ko je Slovenija odprla projekt "1000 mladih raziskovalcev", ki je omogočil vključitev podiplomskih študentov v raziskovalno delo. Po diplomi na Oddelku za fiziko, Fakultete za matematiko in fiziko, Univerze v Ljubljani se je Dragan odločil za magistrski študij na Oddelku za matematiko, raziskovalno delo pa je nadaljevalna Kemijskem inštitutu v moji skupini. Pravočasno je opravil vse potrebno za pridobitev stopnje magistra matematičnih znanosti. Sodeloval je pri znanstvenem delu, ki je bilo objavljeno v desetih člankih v mednarodnih znanstvenih revijah. Dragan je sodeloval pri raziskavah na precej raznolikih področjih, kot so stroga statistična mehanika, uporaba molekularne dinamike za simulacijo bioloških sistemov in celo pri tehnološko usmerjenem študiju mehanskih lastnosti fulerenov.
prof. dr. Branko Borštnik,
dolgoletni vodja laboratorija za molekularno modeliranje na Kemijskem inštitutu v Ljubljani
branko.borstnik@ki.si

Mag. Dragan Lukman, ki je imel magisterij iz matematike ter diplomo iz fizike z Univerze v Ljubljani, me je kontaktiral, na osnovi pomembnih priporočil Prof.dr. Norme Mankoč Borštnik, v maju 2019, z zanimanjem za delovno mesto raziskovalca na mojem raziskovalnem projektu Kvantna lokalizacija v kaotičnih sistemih, ki je bil izvajan na CAMTP - Centru za uporabno matematiko in teoretično fiziko Univerze v Mariboru, in je bil financiran s strani ARRS. Že ob prvem intervjuju je postalo nemudoma jasno, da ima kar široke izkušnje v sodelovanju z različnimi raziskovalnimi skupinami v Sloveniji, predvsem z Normo Mankoč Borštnik in njenimi sodelavci, a tudi z drugimi, s širokim znanjem v fiziki in matematiki ter v računski teoretični fiziki. Zato je bila lahka moja odločitev, da mu ponudim zaposlitev. Tako se je pridružil moji jedrni raziskovalni skupini, ki je del CAMTP in katere člani so tudi dr. Qian Wang, dr. Črt Lozej (moj tedanji doktorand) in dr. Benjamin Batistić (tudi moj nekdanji doktorand, 2015, sedaj podoktorski sodelavec). Naše sodelovanje z Draganom se je pričelo 1. junija 2019. Naš glavni predmet raziskav je bil pojav kvantne ali dinamične lokalizacije v klasičnih kaotičnih sistemih, ena glavnih tem na področju kvantnega kaosa. Natančneje, obširno in poglobljeno smo proučevali lokalizacijske pojave v tako imenovanih limonastih biljardih, ki so posebna družina dvo-dimenzionalnih biljardov z izjemno bogatim vedenjem tako klasično kot kvantno. Leti so pomembni paradigmatični modelski sistemi. Izbor teh biljardov je bil omogočen zahvaljujoč obširnim računom Črta Lozeja v teku njegove doktorske disertacije. Dragan je pričel z delom dokaj navdušeno, in je uporabljal v glavnem softverske programe, ki jih je v dolgih letih razvijal in razvil Benjamin Batistić, in ki jih je v zadnjem času zelo korenito izboljšal in razširil Črt Lozej. Na tej poti smo odkrili veliko novih vznemirljivih rezultatov, ki so izšli iz naših masivnih računov, in Dragan je bil vselej zelo skrben, hiter, odziven in zanesljiv, z dobrim fizikalnim vpogledom. Rezultate je predstavil v kar najkrajšem možnem času, pri čemer je delal vsak dan od zgodnjega jutra do poznega popoldneva, pa tudi čez vikend od doma. Na osnovi rezultatov v okviru sodelovanja z njim smo pripravili štiri pomembne članke, trije od njihso že objavljeni v odličnih revijah (Physical Review E, Physics MDPI, Nonlinear Phenomena in Complex Systems), četrti pa je v procesu pisanja. Zato je Draganov prispevek k našim rezultatom bistven in cenjen.
Dragan je bil prijazna osebnost, zelo skromen in tih, zmerom v pomoč, nikoli se ni pritoževal, ter predan svojemu delu, ne samo na našem institutu, temveč tudi v drugih skupinah. O njegovem zasebnem življenju nismo vedeli veliko, saj je bil zelo plah, in ni kazal čustev, kar pa ne pomeni, da ni bil senzibilen in empatičen.
Tragična novica o njegovi nenadni smrti na njegovi poti na delo zgodaj zjutraj v ponedeljek 19. julija 2021 je bila velik šok za vse nas. Spominjali se ga bomo kot čudovitega kolega in zelo dobrega raziskovalca. Naši skupni članki z njim so trajen spomin nanj.
Prof. dr. Marko Robnik, član EASA
Ustanovitelj in direktor CAMTP - Centra za uporabno matematiko in teoretično fiziko Univerze v Mariboru
Robnik@uni-mb.si

*To Dragan, in grateful memory*

A man of great integrity
A private man,
who shyly would inform you
about things misunderstood, and facts
about cosmology, computers
or Slovenia
His land.
He was a helper, much too humble,
and so gentle
that we sometimes did not see him.
And suddenly he's gone.
A summer day, the brightest day
in early afternoon,
the coffee cup half full, and children
laughing in the park nearby
When suddenly a wind
as light
as butterfly
came by
and brought him
to the other side,
and left us here
in our confusion, our never ending
search for understanding
All our reasoning, our turning
every stone
in this chaotic pain
and beauty
where our lives take place.